%% file: paper.tex
\documentclass[oldversion]
{aa}
\usepackage{amsmath}
\usepackage{natbib}
\usepackage{epsfig}
\usepackage{txfonts}

\input{Befehle}

\newcommand{\change}[1]{{#1}}

\begin{document}

\titlerunning{Cosmological hydrogen recombination: Ly$n$ line feedback and continuum escape} 
\title{Cosmological hydrogen recombination: \\ Ly$n$ line feedback and continuum escape}

\author{J. Chluba\inst{1} \and R.A. Sunyaev\inst{1,2}}
\authorrunning{Chluba \and Sunyaev}

\institute{Max-Planck-Institut f\"ur Astrophysik, Karl-Schwarzschild-Str. 1,
85741 Garching bei M\"unchen, Germany 
\and 
Space Research Institute, Russian Academy of Sciences, Profsoyuznaya 84/32,
117997 Moscow, Russia
}

\offprints{J. Chluba, 
\\ \email{jchluba@mpa-garching.mpg.de}
}

\date{Received 20 February 2007 / Accepted 16 September 2007}

\abstract{We compute the corrections to the cosmological hydrogen
  recombination history due to {\it delayed} feedback of Lyman-series photons
  and the escape in the Lyman-continuum. The former
  process is expected to slightly {\it delay} recombination, while the latter
  should allow the medium to recombine a bit {\it faster}.
  \change{It is shown that} the subsequent feedback of released Lyman-$n$
  photons on the lower lying Lyman-$(n-1)$ transitions yields a maximal
  correction of $\Delta N_{\rm e}/N_{\rm e}\sim 0.22\%$ at $z\sim
  1050$. \change{Including only Lyman-$\beta$ feedback onto the Lyman-$\alpha$
  transition, accounts for most of the effect.}
\change{We find} corrections to the cosmic microwave background $TT$ and $EE$
  power spectra \change{with typical peak to peak amplitude $|\Delta
  C^{TT}_l/C^{TT}_l|\sim 0.15\%$ and $|\Delta C^{EE}_l/C^{EE}_l|\sim 0.36\%$
  at $l\lesssim 3000$}.
The escape in the Lyman-continuum and feedback of Lyman-$\alpha$ photons on
the photoionization rate of the second shell lead to modifications \change{of
the ionization history} which are very small (less than $|\Delta N_{\rm
e}/N_{\rm e}|\sim \text{few}\times 10^{-6}$).  }

\maketitle

\section{Introduction}
\label{sec:Intro}
The tremendous advances in observations of the cosmic microwave background
(CMB) temperature and polarization angular fluctuations
\citep[e.g.][]{Page2006, Hinshaw2006} and the prospects with the {\sc Planck}
Surveyor\footnote{www.rssd.esa.int/Planck} have motivated several groups to
re-examine the problem of cosmological hydrogen recombination,
\change{with the aim of identifying previously neglected physical processes
  which could affect the ionization history of the Universe at the level of
  $\gtrsim 0.1\%$, and may lead to modifications of the cosmological hydrogen
  recombination spectrum
  \citep{Dubrovich1975,Beigman1978,Rybicki93,DubroVlad95,Burgin2003,Dubrovich2004,Kholu2005,Wong2006,Jose2006,
    Chluba2006b, Chluba2007}, which could become observable in the future.}
\change{For example, effects connected with the two-photon transitions from
  high s and d-states to the ground state \citep{Dubrovich2005, Wong2007,
    Chluba2007b}, the induced 2s-two-photon decay \citep{Chluba2006}, the
  increase in the 2s-two-photon absorption rate due to the large
  Lyman-$\alpha$ spectral distortion \citep{Kholu2006}, and details in the
  evolution of the populations of the angular momentum sub-states
  \citep{Jose2006, Chluba2007} were discussed.
All these studies show that sub-percent-level corrections to the ionization
history do exist, which in principle could bias the values of the key
cosmological parameters \citep{Lewis2006}.
It is clear, that when reaching percent-level accuracy in the determination of
the key cosmological parameters or when considering signatures from inflation,
e.g. the possibility of a running spectral index, accurate theoretical
predictions of the ionization history are required.}
Here we examine the effects due to feedback of hydrogen Lyman-series photons
and the escape of photons in the Lyman-continuum on the ionization history
during the epoch of hydrogen recombination\change{, and the impact of this
  process on the Lyman-$\alpha$ distortion of the CMB blackbody spectrum}.

The strongest distortions of the CMB spectrum arising during the epoch of
hydrogen recombination are due to the Lyman-$\alpha$ transition and the 2s
two-photon decay \citep{Zeldovich68, Peebles68}.
The feedback of these excess photons on the photoionization
rates of the second shell has been considered by \citet{SeagerRecfast1999, Seager2000},
with no significant changes for the \change{number density of} free electrons, $N_{\rm e}$.
We confirm this statement for the feedback from Lyman-$\alpha$ photons and
find a maximal correction of $\Delta N_{\rm e}/N_{\rm e}\sim 10^{-6}$ at
$z\sim 500$, the redshift which roughly corresponds to the time, when the
maximum of the CMB spectral distortion due to the Lyman-$\alpha$ transition is
reaching the Balmer-continuum frequency.

However, it has been shown by \citet{Kholu2006} that the huge excess of
photons in the Wien-tail of the CMB due to the Lyman-$\alpha$ distortion leads
to an increase of the ${\rm 1s}\rightarrow{\rm 2s}$ two-photon absorption
rate, which delays recombination and introduces corrections to the ionization
history at the percent-level.
Similarly, one expects some feedback of escaping Lyman-series photons on
lower \change{lying} Lyman-transitions and eventually the Lyman-$\alpha$
transition.
Recently, this process was also considered by \citet{Switzer2007I} for both
helium and hydrogen recombination, using an iterative approach.
Here we directly compute the escaping Lyman-$n$ radiation within the Sobolev
approximation \citep[e.g. see][]{Jose2006, Chluba2007} and include the
feedback on the Lyman-$(n-1)$ transition at all times by evaluating the
distorted spectrum at the corresponding frequency.
In addition, we assume that due to the huge optical depth in all the
Lyman-lines, at the end only the Lyman-$\alpha$ line remains and {\it all the
Lyman-$n$ lines will be completely re-processed} by the closest lower lying
Lyman-$(n-1)$ transition. Therefore, feedback always only works as ${\rm
Ly}n\rightarrow {\rm Ly}(n-1)$.
It is also important to ensure that the line is {\it not} producing any
feedback on itself.

\change{In addition, we estimate} the escape probability for the
Lyman-continuum and include the \change{approximate} net rate for direct
recombinations into our \change{multi-level} code. As argued earlier
\citep{Zeldovich68, Peebles68}, we find that the modification of the
ionization history due to this process is indeed completely negligible.

\begin{table}
\caption{Appearance of the first few Lyman lines for the computations not including any
feedback. In particular we give the approximate redshift, $z_{\rm max}$, at
which $\Delta I_\nu(z_{\rm em})$ is maximal (see Fig. \ref{fig:DI_z}), and the
redshift at which the peak of the line reaches the next lower Lyman resonance,
$z_{\rm f}\approx z_{\rm max}\,\nu_{{\rm Ly}(n-1)}/\nu_{{\rm Ly}n}$, where
$\nu_{{\rm Ly}k}$ is the resonance frequency of the Lyman-$k$ transition. Also
we give the approximate total number of escaping photons per hydrogen nucleus for each transition.}
\label{tab:Lyn}
\centering
\begin{tabular}{@{}lccccc}
\hline
\hline
Line & $n$ & $\nu_{\rm {\rm Ly}n}$ [Hz] & $z_{\rm max}$ & $z_{\rm f}$ &
$N_\gamma/N_{\rm H}$\\
\hline
Lyman-$\alpha$ & 2 & $\pot{2.4674}{15}$ & $1400$ & - & $\pot{4.3}{-1}$ \\
Lyman-$\beta$ & 3 & $\pot{2.9243}{15}$ & $1450$ & $1223$ & $\pot{1.8}{-3}$ \\
Lyman-$\gamma$ & 4 & $\pot{3.0842}{15}$ & $1461$ & $1385$ & $\pot{2.9}{-4}$\\
Lyman-$\delta$ & 5 & $\pot{3.1583}{15}$ & $1464$ & $1430$ & $\pot{1.3}{-4}$ \\
Lyman-$\epsilon$ & 6 & $\pot{3.1985}{15}$ & $1467$ & $1449$ & $\pot{7.9}{-5}$ \\
\hline
\hline
\end{tabular}
\end{table}
\begin{figure}
\centering 
\includegraphics[width=0.98\columnwidth]{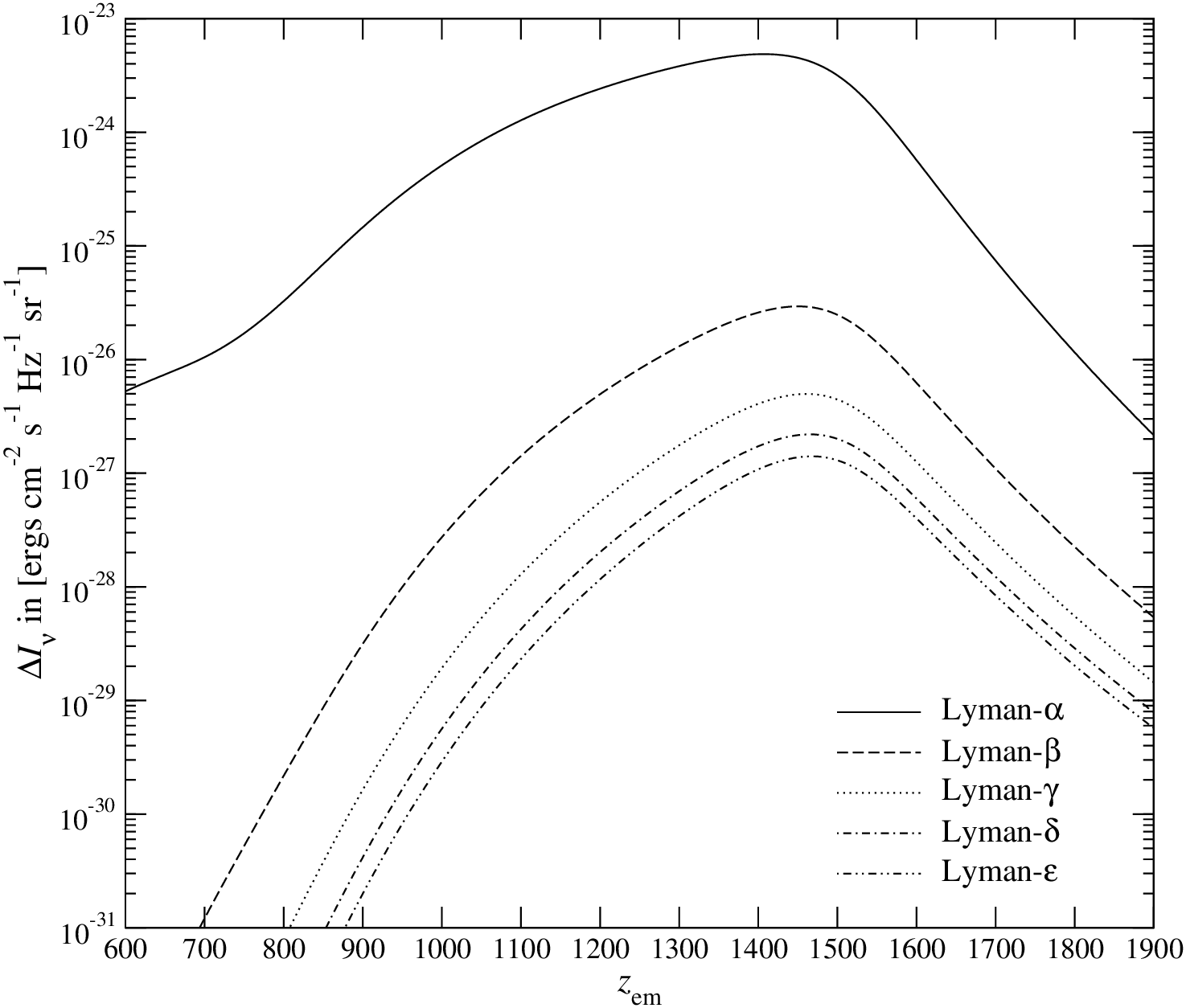}
\\[1mm]
\includegraphics[width=0.98\columnwidth]{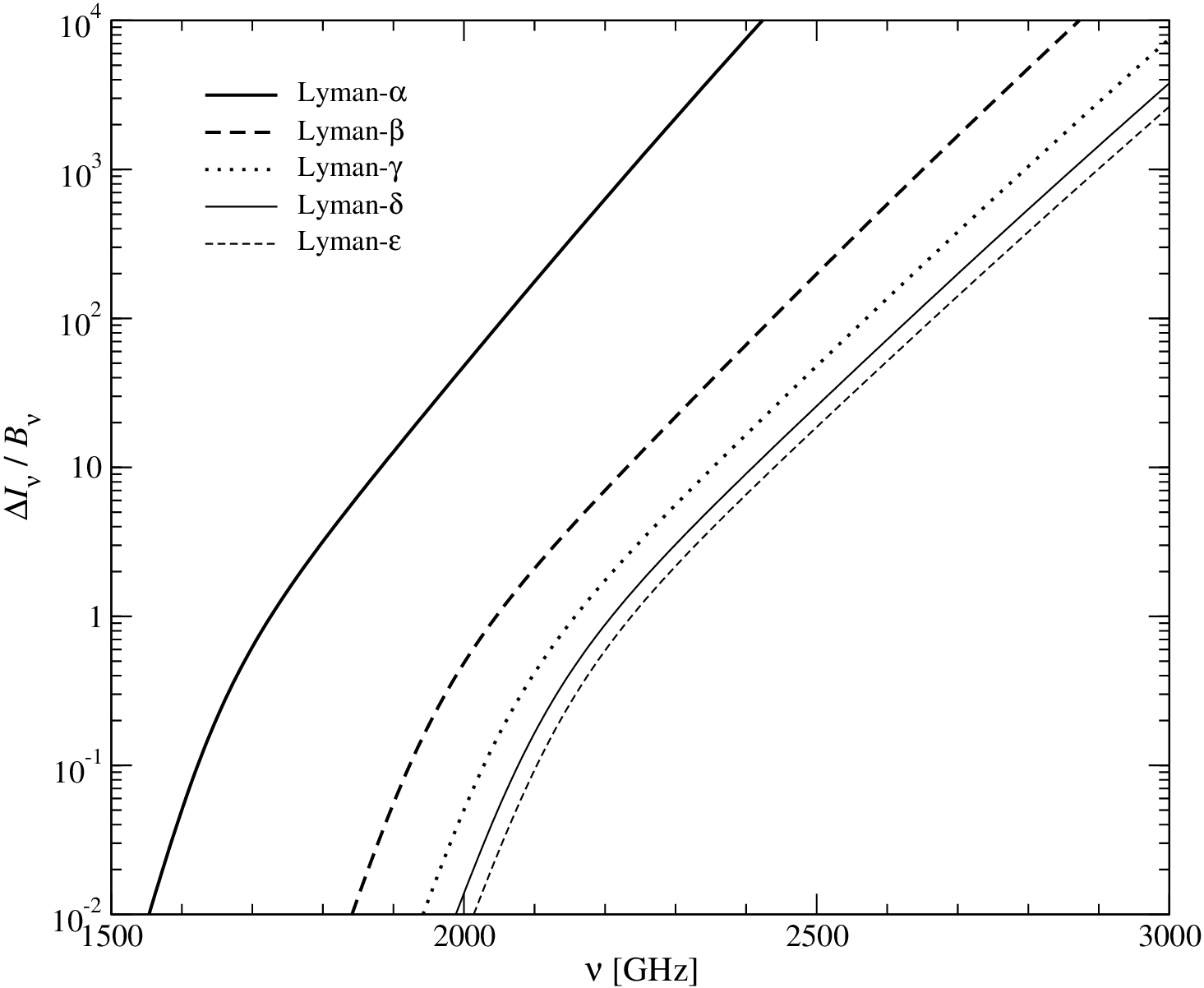}
\caption{Spectral distortions due to the first few Lyman-transitions at $z~=~0$.
  No feedback ${\rm Ly}n\rightarrow {\rm Ly}(n-1)$ has been included.
{\it Upper panel}: as a function of the redshift of emission \citep[see Eq. 1
in][]{Jose2006}.
{\it Lower panel}: relative to the CMB blackbody as a function of observing
frequency $\nu$.  
}
\label{fig:DI_z}
\label{fig:Lyman-series}
\end{figure}
\begin{figure}
\centering 
\includegraphics[width=0.96\columnwidth]
{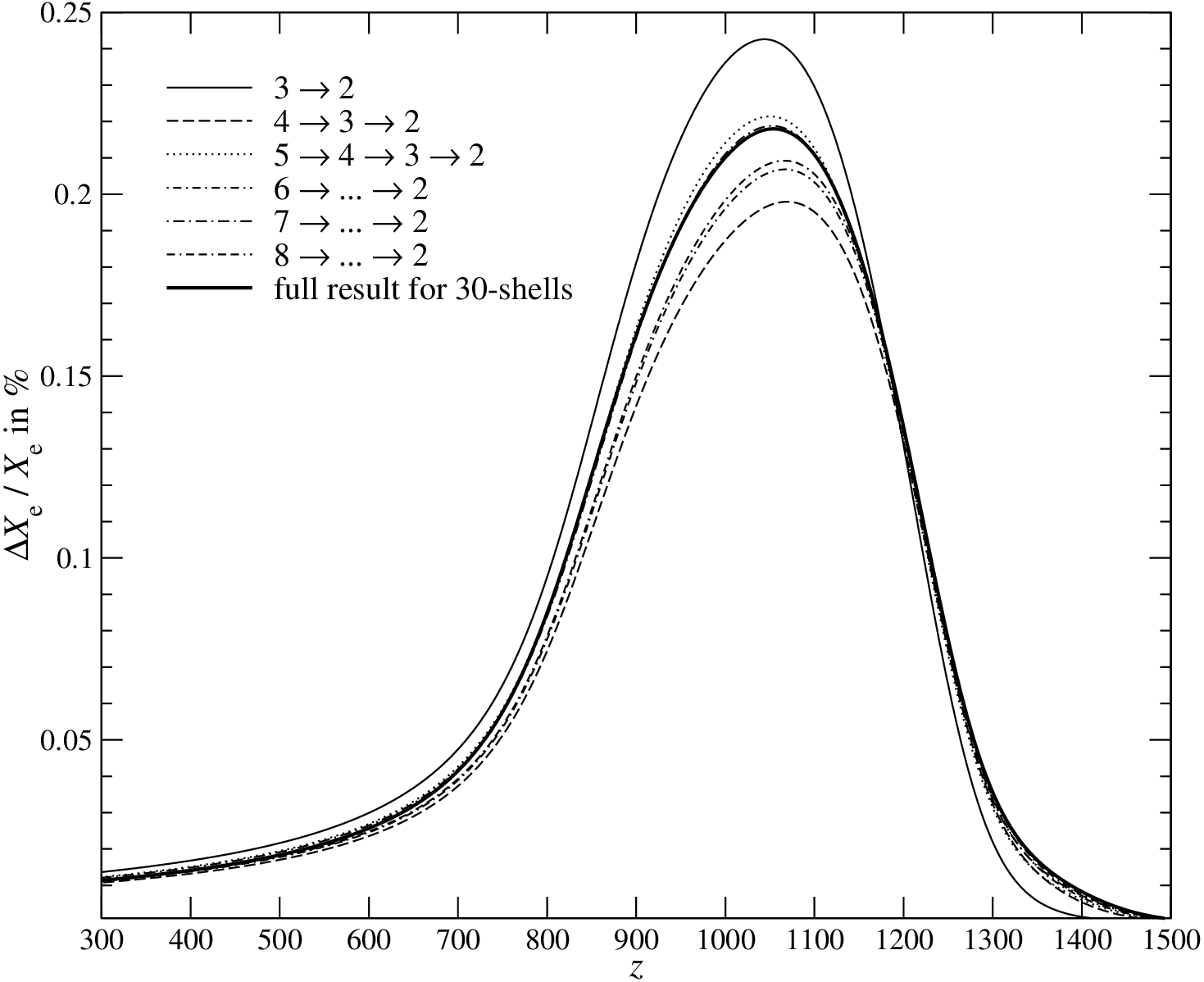}
\caption
{Changes in the free electron fraction due to inclusion of Lyman-series
feedback relative to the reference model without feedback.
The computations were performed for a 15-shell atom where for the Lyman-series
the escape of photons was modeled using the Sobolev approximation. The curves
are labeled according to the sequence of feedback that was included. For
example, $4\rightarrow3\rightarrow2$ indicates that Ly$\gamma\rightarrow$
Ly$\beta$ and Ly$\beta\rightarrow$ Ly$\alpha$ feedback was taken into account.
Also the full result for the 30-shell hydrogen atom is shown, where feedback
up to $n=30$ was included.  }
\label{fig:Lyn}
\end{figure}
\section{Re-absorption of Lyman-series photons}
\label{sec:Lyman-series}
Although in computations with no feedback, the contributions of the $n{\rm
  p}\rightarrow {\rm 1s}$ transitions from shells with $n>2$ to the total
Lyman-series are rather small \citep[see Fig.~8 in][]{Jose2006}, in the
Wien-tail of the CMB the intensity of these lines is significantly larger than
the blackbody (see Fig.~\ref{fig:Lyman-series}).
Therefore one expects that after redshifting each Lyman-transition from $n>2$
will increase the absorption rate for the closest lower Lyman-series
transition ($n'=n-1\geq 2$) and thereby slightly {\it delay} hydrogen
recombination.
For example, the Lyman-$\beta$ transition peaks at $z_{\rm em}\sim 1450$ with
contribution $\Delta I_\nu/B_\nu\sim 64\%$ relative to the CMB blackbody
spectrum $B_\nu$. These excess photons reach the central frequency of the
Lyman-$\alpha$ transition at $z\sim 1223$ and increase the Lyman-$\alpha$
absorption rate, thereby delaying recombination (see Table \ref{tab:Lyn}).
It is easy to estimate the total number of photons released in the
Lyman-$\beta$ transition using Fig.~\ref{fig:Lyman-series}. Its value is close
to $\sim 0.18\%$ of the total number of hydrogen nuclei (see Table
\ref{tab:Lyn} for the other transitions).
At the time of maximal feedback ($z_{\rm f}\sim 1223$) the number of free
electrons is roughly $38\%$ of the total number of hydrogen nuclei.
Therefore the maximal effect one can expect for the feedback-induced
corrections to the ionization history from the Lyman-$\beta$ line is $\Delta
N_{\rm e}/N_{\rm e}\lesssim 0.5\%$.

In Fig. \ref{fig:Lyn} we present the results for detailed computations of
\change{feedback-induced} changes in the free electron fraction. Including
only Lyman-$\beta$ feedback on the Lyman-$\alpha$ transition accounts for most
of the effect, leading to a maximal difference of $\Delta N_{\rm e}/N_{\rm
  e}\sim 0.25\%$ at $z\sim 1040$.

When including in addition the feedback of Lyman-$\gamma$ on Lyman-$\beta$, the
maximal relative difference decreases to $\Delta N_{\rm e}/N_{\rm e}\sim
0.20\%$, now peaking at $z\sim 1070$. 
This behavior is expected, since the Lyman-$\gamma$ photon reduces the
transition rate in the Lyman-$\beta$ channel such that its feedback on the
Lyman-$\alpha$ line should become slightly smaller. In addition, the number of
electrons reaching the ground state via the Lyman-$\beta$ transition will be
reduced, which very likely leads to the small increase of the correction at
$z\gtrsim 1200$.

\begin{figure}
\centering 
\includegraphics[width=0.93\columnwidth]
{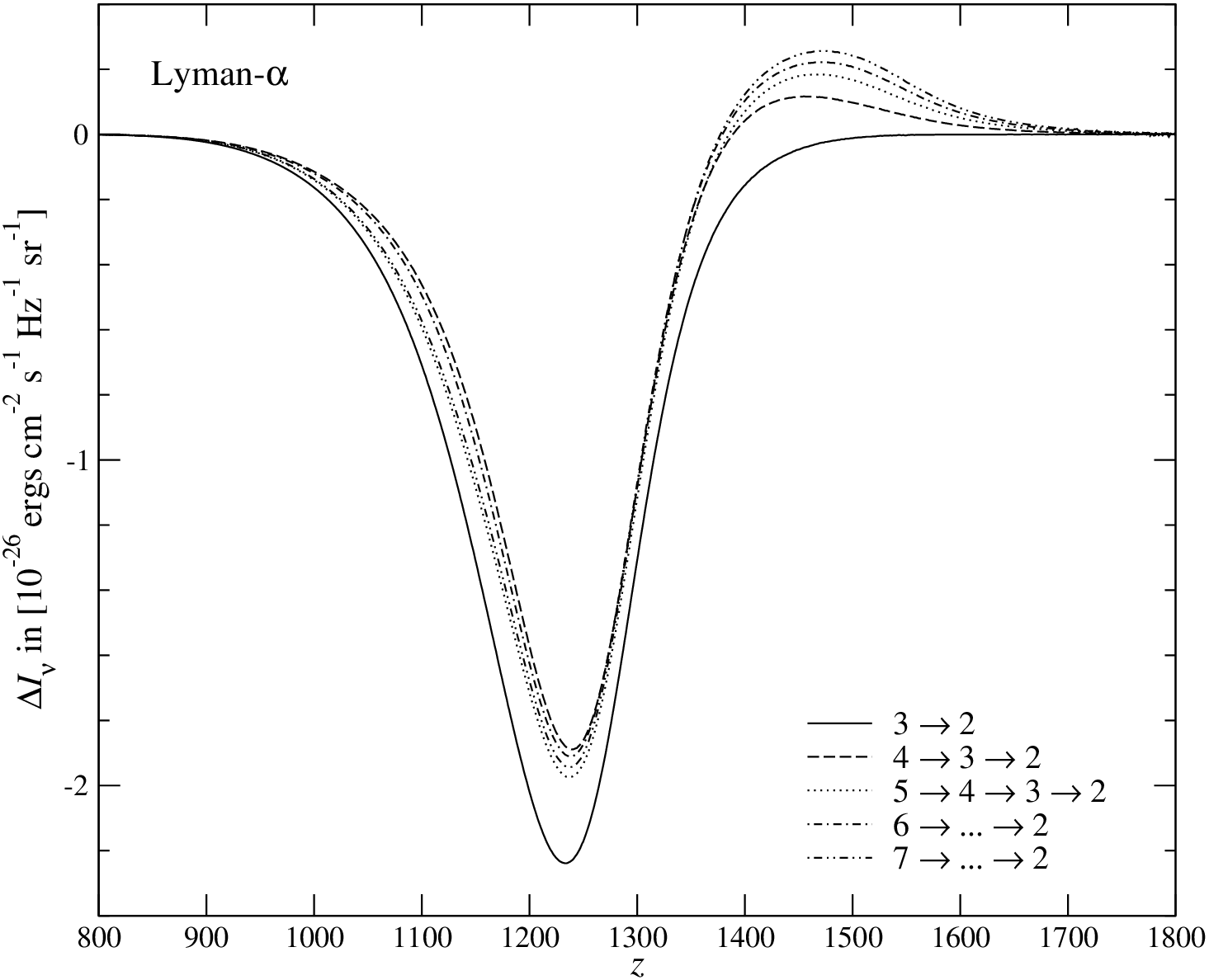}
\\
\includegraphics[width=0.93\columnwidth]
{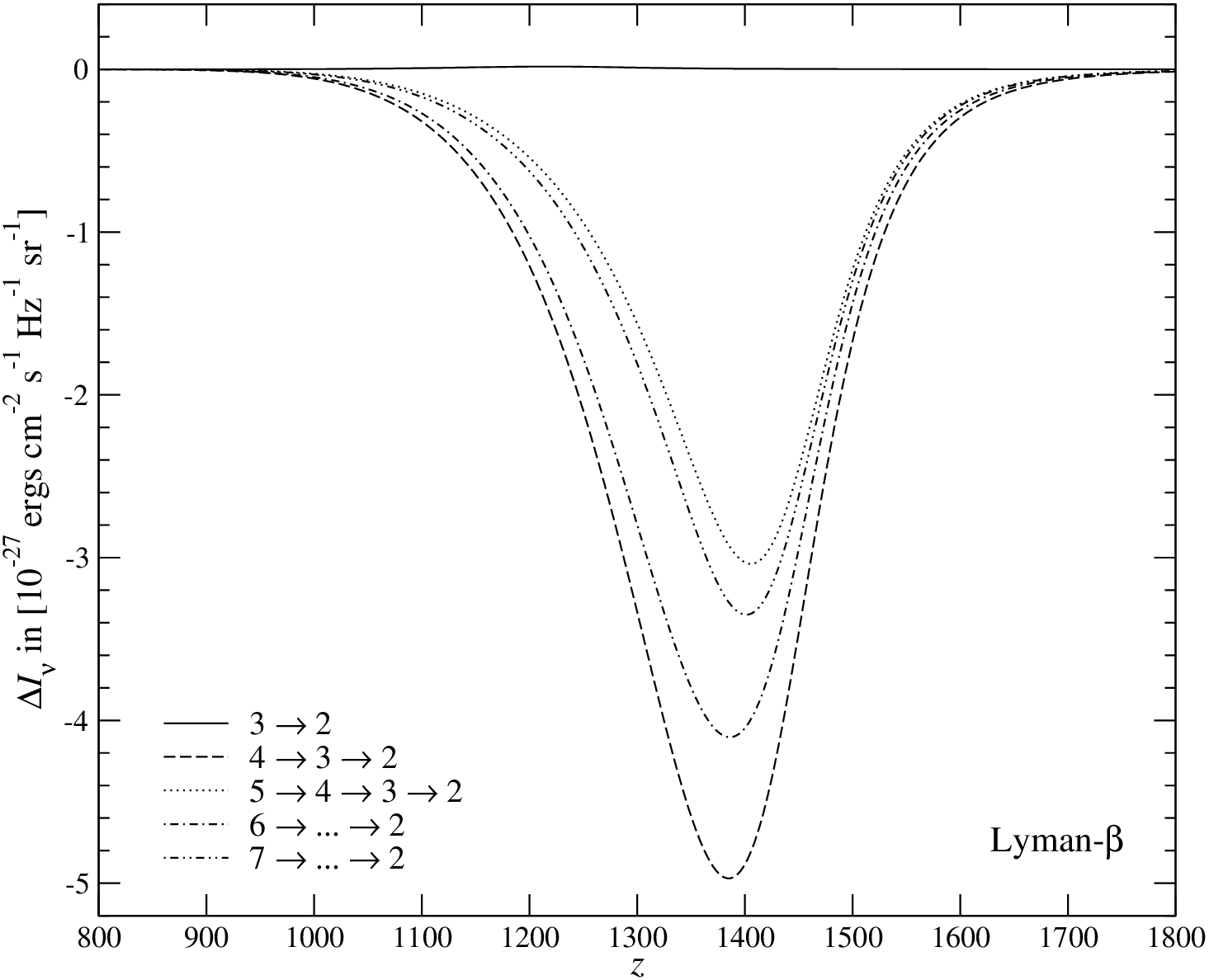}
\\
\includegraphics[width=0.93\columnwidth]
{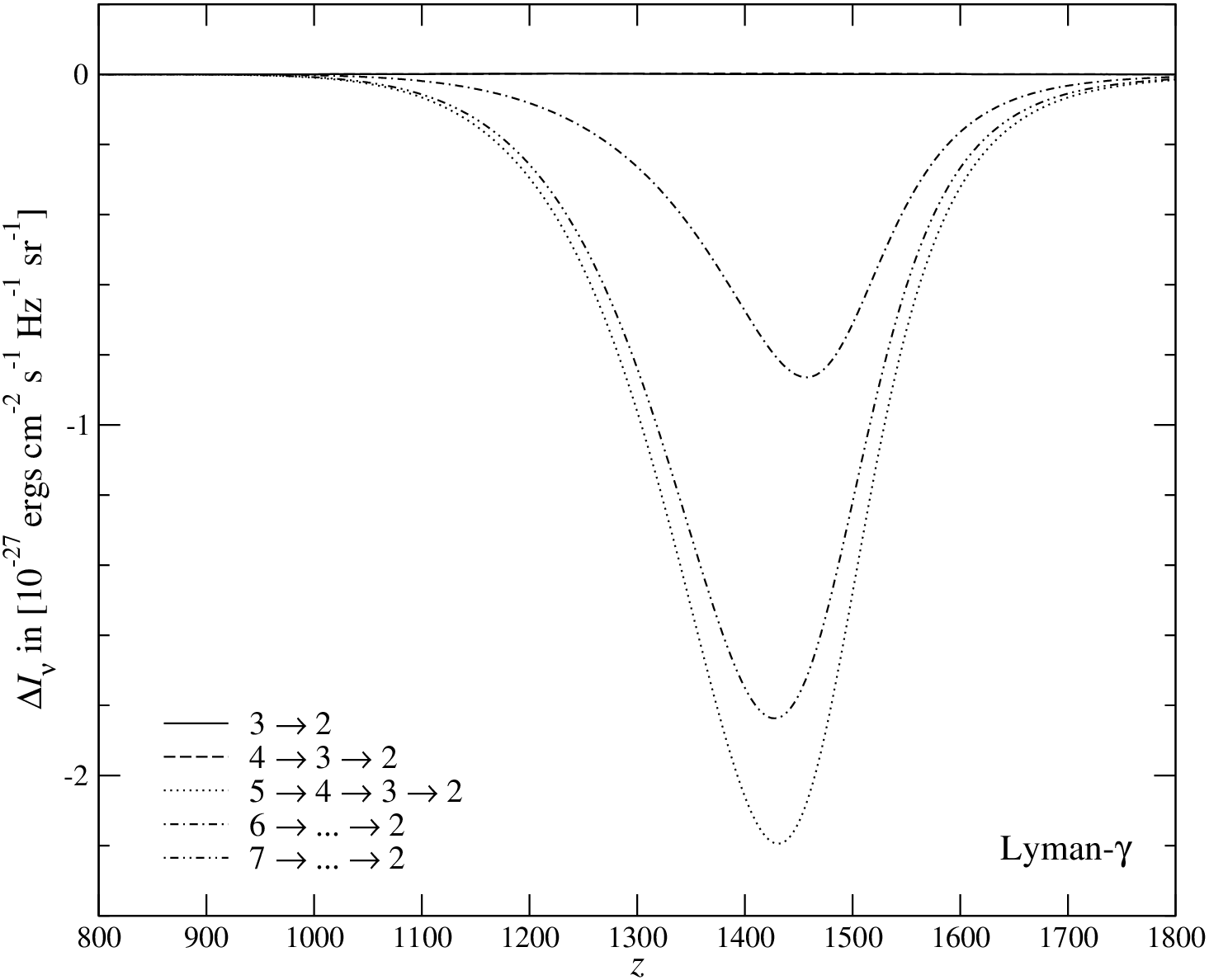}
\caption
{Feedback-induced changes of $\Delta I_\nu(z_{\rm em})$ for the Lyman-$\alpha$
  and the modifications of the Lyman-$\beta$ and $\gamma$ lines {\it before}
  their absorption within the corresponding lower lying resonance.
  The results are based on the computations for a 15-shell hydrogen atom.}
\label{fig:DI.Ly2}
\end{figure}

\begin{figure}
\centering 
\includegraphics[width=0.96\columnwidth]
{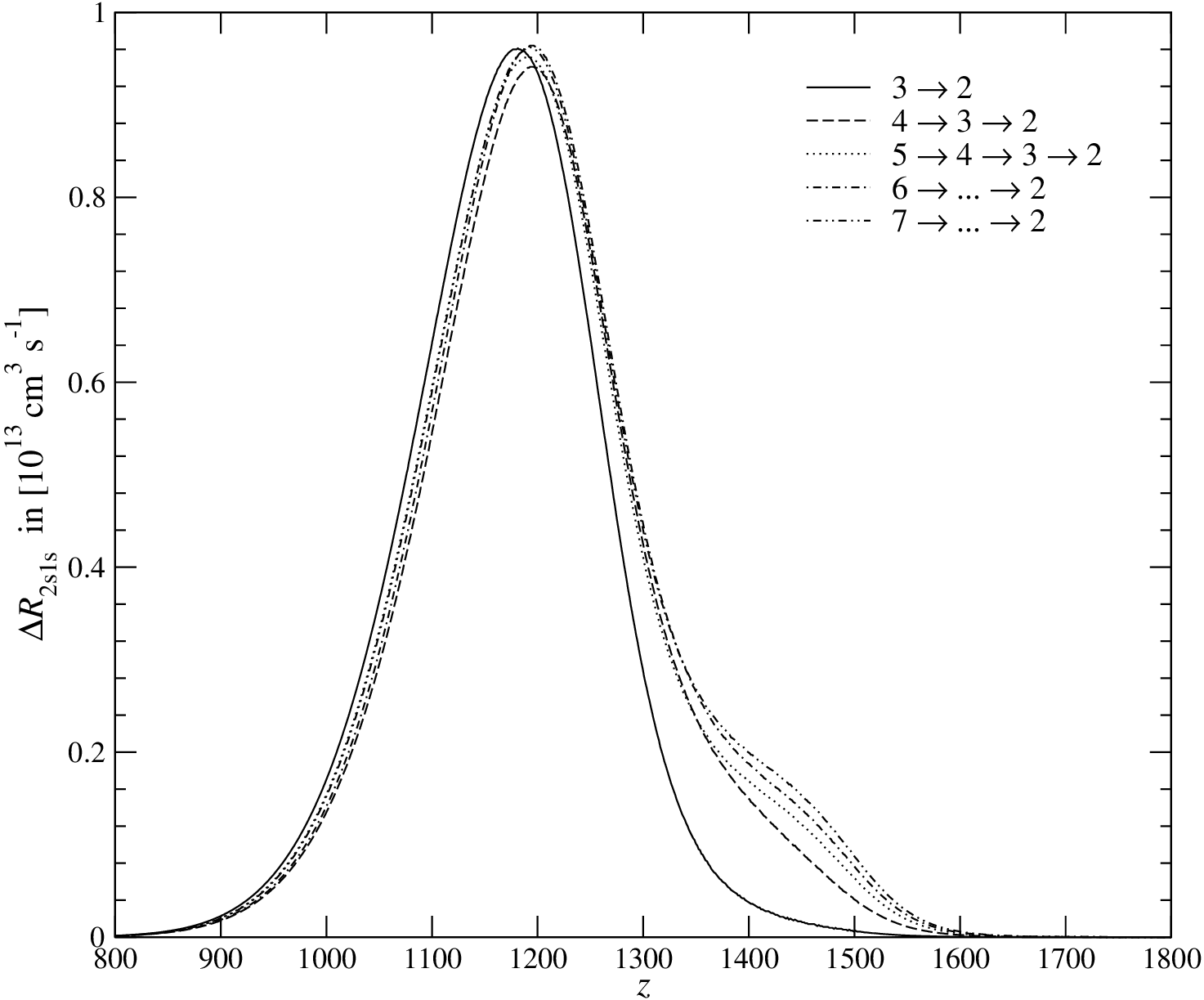}
\caption
{Changes in the net 2s-1s two-photon decay rate due to the inclusion of
${\rm Ly}n\rightarrow {\rm Ly}(n-1)$ feedback, as labeled. The
results are based on the computation for a 15-shell hydrogen atom.}
\label{fig:DR.2s}
\end{figure}

Subsequent inclusion of more feedback results in an alternating behavior of
$\Delta N_{\rm e}/N_{\rm e}$. 
We found that for 30-shells, convergence is reached when including feedback
from $n$p-levels with $n\lesssim 15$. In Fig. \ref{fig:Lyn} we also give the
full result including {\it all} feedback within a 30-shell hydrogen atom.
In addition we found that the correction barely depends on the total number of
included shells.
Therefore we expect that even for computations with up to 100-shells the total
feedback-induced relative difference will not change significantly.
Examining the final \change{differences} in the Lyman-$\alpha$ distortion and
the feedback-induced modifications of the Lyman-$\beta$ and $\gamma$ lines
{\it before} their absorption within the corresponding lower lying resonance
(see Fig.  \ref{fig:DI.Ly2}) shows that the situation is a bit more
complicated.
Including only Lyman-$\beta$ feedback the amount of emission in the
Lyman-$\alpha$ line reduces by $\sim 0.8\%$ at $z\sim 1232$. Note that this
redshift is very close to the expected value, $z_{\rm f}\sim 1223$, for the
maximal feedback (see Table \ref{tab:Lyn}).
On the other hand, due to the feedback-induced small changes in the
populations of the levels, the net 2s-1s two-photon transition rate also
increases with a maximum of roughly $\sim 0.5\%$ at $z\sim 1034$ (see Fig.
\ref{fig:DR.2s}), which partially cancels the delaying feedback effect on the
Lyman-$\alpha$ transition.

When including Lyman-$\gamma$ feedback, as expected, the emission in the
Lyman-$\beta$ line reduces, with a maximal difference at the expected value
$z\sim 1384$. Adding Lyman-$\delta$ feedback reduces the strength of the
feedback of Lyman-$\gamma$ on the Lyman-$\beta$ transition by a significant
fraction. Comparing the relative strength of the Lyman-$\delta$ at its maximum
with the Lyman-$\gamma$ line at $z_{\rm f}$ (Fig.~\ref{fig:DI_z}) yields
$\Delta I^\delta_\nu(z_{\rm max})/\Delta I^\gamma_\nu(z_{\rm f})\sim 1/2$,
which indicates that Lyman-$\delta$ should be able to affect Lyman-$\gamma$
strongly.
Looking at the relative difference in the Lyman-$\gamma$ when including
Lyman-$\delta$ feedback shows that at $z_{\rm f}\sim 1430$ the line is
reduced by $\sim 50\%$. 

\section{The Lyman-continuum escape}
\label{sec:Lyc_esc}
To estimate the maximal effect arising due to {\it successful} escape
of photons from the Lyman-continuum we shall neglect any re-absorption of
photons below the threshold frequency $\nu_{\rm c}\approx
\pot{3.288}{15}\,$Hz. Although one can expect some modification due to the
{\it forest} of lines in the Lyman-series close to the
continuum\footnote{Within a small energy-distance $\xi$ below the ionization
  potential these in principle can be treated as a {\it continuation} of the
  photoionization cross-section \citep[e.g. see Appendix II
  in][]{Beigman1968}.}, this approach will yield an upper limit on the effect,
which turns out to be very small in any case.

With this in mind, in the vicinity of the Lyman-continuum the photon number
only changes by recombination and ionization to the ground state (we neglect
electron scattering):
\begin{equation}
\label{eq:DNLyc}
\frac{1}{c}\!\left.\pAb{N_\nu}{t}\right|^{\rm rec}_{\rm 1s} 
\!\!=N_{\rm e} N_{\rm p}\,f(\Te)\,
\sigma_{{\rm 1sc}} 
\frac{2 \nu^2}{c^2} e^{-\frac{h\nu}{\kB\Te}}
-N_{\rm 1s}\,\sigma_{{\rm 1s c}} N_\nu 
\Abst{.}
\end{equation}
Here $N_{\rm e}$ and $N_{\rm p}$ are the free electron and proton number densities,
$N_{\rm 1s}$ is the number density of hydrogen atoms in the ground state and
$N_\nu=I_\nu/h\nu$, where $I_\nu$ is the photon intensity. In addition,
$f(\Te)=\left(N_{\rm 1s}/N_{\rm e}\,N_{\rm p}\right)^{\rm LTE}$
and $\sigma_{{\rm 1s c}}$ is the 1s-photoionization cross-section.
As can be seen from Eq. \eqref{eq:DNLyc}, photons are emitted with a
spectrum\footnote{Here and below we neglect the difference in the electron and
  photon temperature, which is smaller than $\Delta T/T\sim 10^{-6}-10^{-4}$
  at most times during recombination.}
$\phi^{\rm rec}_{\rm em}\propto \sigma_{{\rm 1s c}}(\nu)\,\nu^2
e^{-h\nu/\kB\Tg}$.
We now compute the probability that {\it one} single recombination will lead
to a successful escape of photons.

The optical depth for absorption of a photon, which has been emitted at
redshift $z_{\rm em}$ and is observed at frequency $\nu_{\rm obs}$ at redshift
$z_{\rm obs}$, within the Lyman-continuum is given by
\beal
\label{app:tau_c}
\tau_{\rm c}(z_{\rm em}, z_{\rm obs}, \nu_{\rm obs})
=c\int^{z_{\rm em}}_{z_{\rm obs}} \frac{N_{1s}\,\sigma_{{\rm 1s c}}(\nu_z)}{H\,(1+z)} \id z
\end{align}
where $\nu_z=\nu_{\rm obs}(1+z)/(1+z_{\rm obs})$. 
If we substitute $\tilde{\nu}=\nu_z$ and use $\id\tilde{\nu}=\nu_{\rm
obs}\id z/(1+z_{\rm obs})$ then we have
\beal
\label{app:tau_c_nuem}
\tau_{\rm c}(z_{\rm em}, \nu_{\rm em}, \nu_{\rm obs})
=c\int^{\nu_{\rm em}}_{\nu_{\rm obs}} \frac{N_{1s}\,\sigma_{{\rm 1s c}}(\tilde{\nu})}{H} \frac{\id \tilde{\nu}}{\tilde{\nu}}
\end{align}
Fixing the emission frequency and the emission redshift one can immediately
determine the current redshift as a function of $\tilde{\nu}$ with
$\tilde{\nu}/\nu_{\rm em}=(1+z)/(1+z_{\rm em})$. 
Now the escape probability in the Lyman-continuum can be directly given in
terms of the normalized emission profile $\phi^{\rm rec}_{\rm
em}=\mathcal{N}\sigma_{{\rm 1s c}}(\nu)\,\nu^2 e^{-h\nu/\kB\Tg}$:
\beal
\label{app:p_S_c}
P^{\rm Ly-c}_{\rm esc}(z_{\rm em})
=\int_{\nu_{\rm c}}^\infty \phi^{\rm rec}_{\rm em}(\nu')
\exp\left[-\tau_{\rm c}(z_{\rm em}, \nu', \nu_{\rm c})\right]\id\nu',
\end{align}
where $\mathcal{N}$ is determined from the condition $\int \phi^{\rm rec}_{\rm
em}(\nu)\id\nu =1$.
Because $\phi^{\rm rec}_{\rm em}$ is sufficiently narrow ($\Delta \nu/\nu_{\rm
  c} \sim \kB \Tg/ h \nu_{\rm c}\sim 1/40$ at $z\sim 1400$) it only takes a very
short time to cross the emission profile via redshifting. Since the
characteristic time for changes of $N_{1s}$ and $H$ is much longer, one can
write
\beal
\label{app:p_S_c_approx}
P^{\rm Ly-c}_{\rm esc}(z_{\rm em})
\approx
\int_{\nu_{\rm c}}^\infty \phi^{\rm rec}_{\rm em}(\nu')
\exp\left[-\tau_0
\int_{\nu_{\rm c}}^{\nu'}\frac{\hat{\sigma}_{{\rm 1sc}}(\tilde{\nu})}{\tilde{\nu}}\id \tilde{\nu}
\right]\id\nu'
\end{align}
where $\hat{\sigma}_{{\rm 1s c}}(\nu)=\sigma_{{\rm 1s c}}(\nu)/\sigma_{{\rm 1s
c}}(\nu_{\rm c})$ and the optical depth at the threshold frequency is
$\tau_0(z_{\rm em})=c \,\sigma_{{\rm 1s c}}(\nu_{\rm c})\,N_{1s}(z_{\rm
em})/H(z_{\rm em})$.

\begin{figure}
\centering 
\includegraphics[width=0.96\columnwidth]
{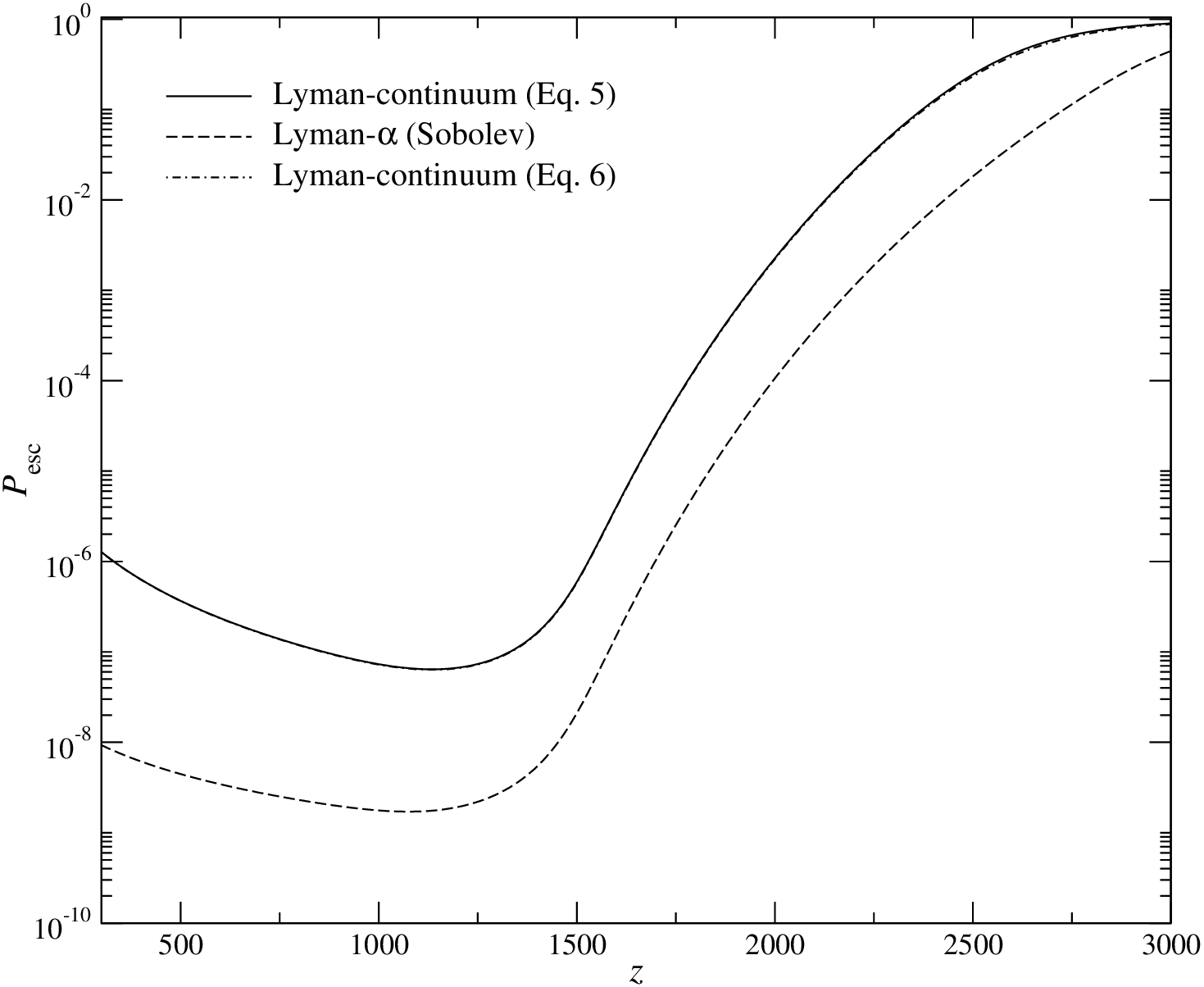}
\caption
{Escape probability for the Lyman-continuum as given by
  Eq.~\eqref{app:p_S_c_approx} and Eq.~\eqref{app:p_S_c_approx_b}. For
  comparison we also show the Sobolev escape probability for the
  Lyman-$\alpha$ photons.}
\label{fig:one}
\end{figure}

It is possible to simplify Eq. \eqref{app:p_S_c_approx} even further. First,
since the emission profile is very narrow one may use $\phi^{\rm rec}_{\rm
  em}\approx \tilde{\mathcal{N}} e^{-h[\nu-\nu_{\rm c}]/\kB\Tg}$, with the
corresponding normalization constant $\tilde{\mathcal{N}}$. Assuming
$\hat{\sigma}_{\rm 1sc}\approx [\nu_{\rm c}/\nu]^3$ one then obtains
\beal
\label{app:p_S_c_approx_b}
P^{\rm Ly-c}_{\rm esc}(z_{\rm em})\approx \int_0^\infty e^{-\xi[1+\tau^{\rm esc}_{\rm c}]}
\id\xi=\frac{1}{1+\tau^{\rm esc}_{\rm c}},
\end{align}
%
where $\tau^{\rm esc}_{\rm c}=\tau_0 \frac{\kB\Te}{h\nu_{\rm c}}$. 
We see that \change{interpreting the continuum profiles as a very} narrow
line\change{\footnote{This is possible due to the exponential factor in
$\phi^{\rm rec}_{\rm em}$.}}, $P^{\rm Ly-c}_{\rm esc}$ \change{appears to be}
similar to the Sobolev approximation for the escape of photons from resonance
lines.
\change{In particular $P^{\rm Ly-c}_{\rm esc}$ also scales as $1/\tau^{\rm
esc}_{\rm c}$ for large optical depth $\tau^{\rm esc}_{\rm c}\gg 1$. However,
unlike in the Sobolev approximation there is {\it no} symmetry between the
absorption and emission profile, which actually is one of the key requirements
that makes the Sobolev escape probability completely independent of the line
profile. }
Comparing the results of Eq.~\eqref{app:p_S_c_approx} and
Eq.~\eqref{app:p_S_c_approx_b} shows that the latter equation provides a
sufficient approximation to $P^{\rm Ly-c}_{\rm esc}$ (see Fig.
\ref{fig:one}).

\begin{figure}
\centering 
\includegraphics[width=0.96\columnwidth]
{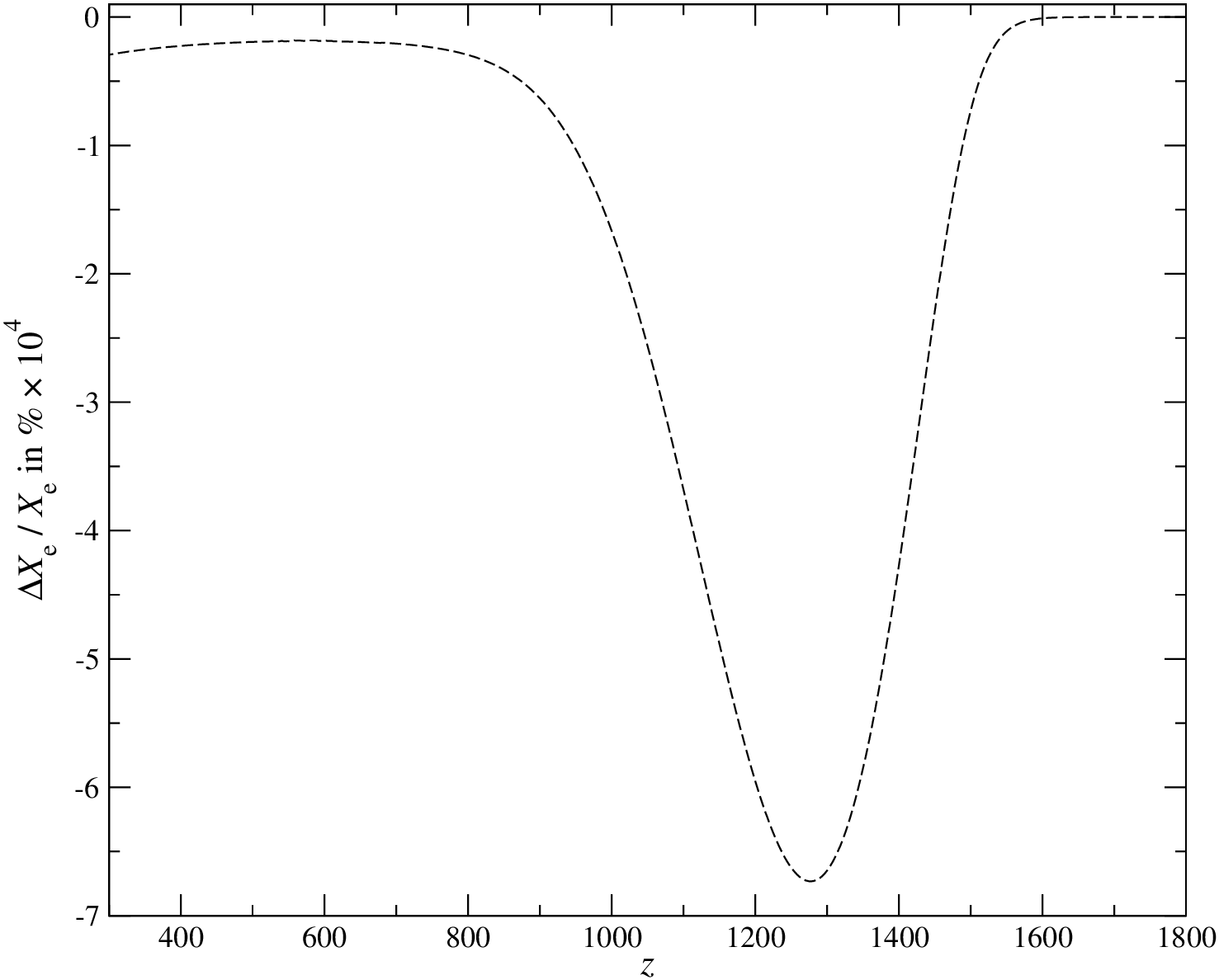}
\caption
{Changes in the free electron fraction due to inclusion of Lyman-continuum
escape relative to the reference model without any direct recombinations to
1s. The computations were performed for a 5-shell atom where for the
Lyman-series escape the Sobolev approximation was used. Note that the absolute
value of $\Delta X_{\rm e}/X_{\rm e}$ is $\sim 10^{-6}$ only.}
\label{fig:two}
\end{figure}

Using Eq.~\eqref{app:p_S_c_approx_b} it is easy to include the possibility of
direct recombinations to the ground state connected with the escape of
continuum photons. For this,one should add
\beal
\label{app:DR_approx}
\Delta R_{\rm c1s}\approx P^{\rm Ly-c}_{\rm esc}(z_{\rm em})\times
\left[N_{\rm 1s}\,R_{{\rm 1sc}}-N_{\rm e}\,N_{\rm p}\,R_{{\rm c1s}}\right],
\end{align}
to the rate equations of the electrons and subtract this term for the
1s-equation. Here $R_{{\rm c1s}}$ and $R_{{\rm 1sc}}$ are the recombination
and photoionization rate of the ground state, respectively.
Note that we assume that on the {\it blue} side of the Lyman-continuum the
spectrum is Planckian. Therefore the photoionization rate $R_{{\rm 1sc}}$ does
not include any spectral distortion. However, due to the release of photons
during $\ion{He}{i}$ recombination some small distortions are created, which
lead to an increase of the photoionization rate.

Comparing the escape probability for the Lyman-continuum with the Sobolev
escape probability for the Lyman-$\alpha$ photons (see Fig. \ref{fig:one}) one
can see that $P^{\rm Ly-c}_{\rm esc}(z_{\rm em})$ is roughly $10-100$ times
larger at most times.
However, including the possibility of direct recombinations to the ground
state in our recombination code yields a tiny correction to $X_{\rm e}$ (see
Fig.~\ref{fig:two}), which shows that $\Delta R_{\rm c1s}$ as given by
Eq.~\eqref{app:DR_approx} is still many orders of magnitude smaller than the net
Lyman-$\alpha$ transition rate. One can safely neglect direct recombinations
to the ground state of hydrogen for computations of the recombination history.

\begin{figure}
\centering 
\includegraphics[width=0.96\columnwidth]
{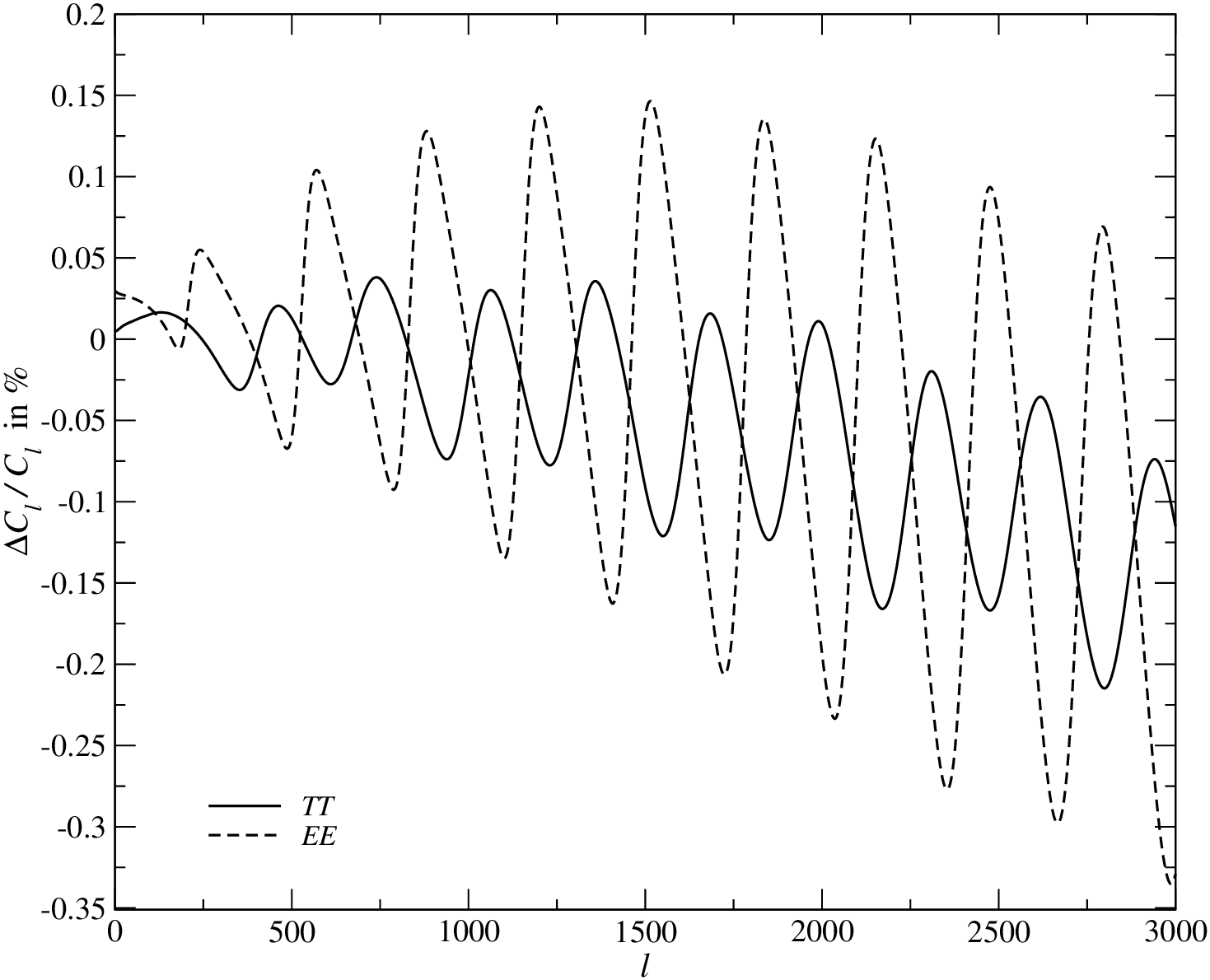}
\caption
{Changes in the CMB temperature ($TT$) and polarization ($EE$) power spectra.
  The differences were computed using our modified versions of \change{{\sc
      Cmbeasy} \citep{Doran2005}, which allows loading of pre-calculated
    recombination histories, and where the corresponding {\sc Recfast}-routine
    was improved to achieve higher numerical accuracy with solvers from the
    {\sc Nag}-library}.}
\label{fig:Dcl}
\end{figure}

\section{Discussion}
\label{sec:discussion}
We have shown that ${\rm Ly}n\rightarrow {\rm Ly}(n-1)$ feedback leads to a
modification in the ionization history of $\Delta N_{\rm e}/N_{\rm e}\sim
0.22\%$ at $z\sim 1050$.
\change{Since this is very close to the peak of the Thomson-visibility
function \citep{Sunyaev1970}, corrections to the CMB power spectra of similar
order are expected. Computing the CMB $TT$ and $EE$
power spectra using the results for the ionization history with and without
the additional feedback we find changes with a typical peak to peak
amplitude $|\Delta C^{TT}_l/C^{TT}_l|\sim 0.15\%$ and $|\Delta
C^{EE}_l/C^{EE}_l|\sim 0.36\%$ at $l\lesssim 3000$ (see Fig. \ref{fig:Dcl})}.

\change{As shown by \citet{Chluba2007b} due to two-photon processes
  one can expect some asymmetries in the emission profiles of the higher
  Lyman-series lines. This could modify the amount and time-dependence for the
  ${\rm Ly}n\rightarrow {\rm Ly}(n-1)$ feedback process and may lead to
  additional differences in the results presented above. However, because the
  bulk of photons that escape in the red wing of a particular Lyman-transition
  is expected to come from close to the line center, this may be of minor
  importance. Here the most interesting aspect may be the changes in the
  time-dependence of the feedback, but more detailed computations would be
  necessary to understand this problem.}

In addition one should look at the feedback-induced corrections \change{in the
  hydrogen recombination history} due to the $\ion{He}{i}$ spectral
distortions, in particular due \change{to the photons appearing in the
  $2^1{\rm P}_1 \rightarrow 1^1 {\rm S}_0$ resonance transition, the $2^3{\rm
    P}_1 \rightarrow 1^1 {\rm S}_0$ intercombination line and the $2^1{\rm
    S}_0 \rightarrow 1^1 {\rm S}_0$ two-photon continuum}.
It is clear that all high frequency $\ion{He}{ii}$ lines will be re-processed
during $\ion{He}{i}$ recombination, and \change{correspondingly can only
  affect the ionization history} during that epoch.
However, for the feedback of the $\ion{He}{i}$ lines on hydrogen recombination
it will be important to compute the re-processing \change{of photons} in the
Lyman-continuum and all the subsequent Lyman-series transitions. Since {\it
  all} of these transitions are very optically thick a huge part of the
\change{photons released during} $\ion{He}{i}$ and $\ion{He}{ii}$
recombination will never \change{reach the observer today, but due to
  re-absorption by neutral hydrogen, they are fully converted to hydrogen
  Lyman-$\alpha$ and 2s-1s continuum photons \citep[see
  also][]{Kholupenko2007, Jose2007}.
Since most of the helium $2^1{\rm P}_1 \rightarrow 1^1 {\rm S}_0$ photons are
released at $z\sim 1800-2600$, they should be re-absorbed by neutral hydrogen
atoms at the early stages $z\gtrsim 1200-1400$.
One can also conclude this from the paper of \citet{Kholupenko2007}, where the
re-processed $\ion{He}{ii}\rightarrow\ion{He}{i}$ photons appear on the red
side of the hydrogen Lyman-$\alpha$ distortion (most hydrogen Lyman-$\alpha$
photons are released at $z\sim 1400$). Therefore the helium photons affect the
hydrogen recombination history well before the maximum of the
Thomson-visibility function ($z\sim 1100$), and hence should have a rather
small impact on the CMB power spectra. However, a more careful computation is
required and will be described in a forthcoming paper \citep{Jose2007}.}

\acknowledgements{The authors thank the anonymous referee for his useful
comments.}

\bibliographystyle{aa} 
\bibliography{Lit}

\end{document}

%% file: Befehle.tex
\newcommand{\id}{{\,\rm d}}

\newcommand{\beq}{\begin{equation}}   %

\newcommand{\eeq}{\end{equation}}   %

\newcommand{\beqa}{\begin{eqnarray}}   %

\newcommand{\eeqa}{\end{eqnarray}}   %

\newcommand{\beal}{\begin{align}}

\newcommand{\enal}{\end{align}}

\newcommand{\bspl}{\begin{split}}

\newcommand{\espl}{\end{split}}

\newcommand{\bsub}{\begin{subequations}}

\newcommand{\esub}{\end{subequations}}

\newcommand{\bmulti}{\begin{multline}}   %

\newcommand{\beqm}{\begin{mathletters}}   %

\newcommand{\eeqm}{\end{mathletters}}   %

\newcommand{\Abst}[1]{\,#1}

\newcommand{\kB}{k_{\rm B}}

\newcommand{\Te}{T_{\rm e}}

\newcommand{\Tg}{T_{\gamma}}

\newcommand{\pd}{\partial}

\newcommand{\pAb}[2]{\frac{\displaystyle\pd #1}{\displaystyle\pd #2}}

\newcommand{\pot}[2]{#1 \times 10^{#2}}

